\documentclass[a4paper,11pt]{article}

\pdfoutput=1 

\usepackage{jheppub} 

\usepackage[T1]{fontenc} 

\begin{document}

\title{Testable constraint on near-tribimaximal neutrino mixing}

\author[a]{Biswajoy Brahmachari}

\author[b,c]{Probir Roy}

\affiliation[a]{
Department of Physics, Vidyasagar Evening College, \\
39 Sankar Ghosh Lane, Kolkata 700006,  India}

\affiliation[b]{
 Saha Institute of Nuclear Physics, \\
Kolkata 700064, India}      

\affiliation[c]{
Center for Astroparticle Physics and Space Science, \\
Bose Institute, Kolkata 700091, India}

\emailAdd{biswa.brahmac@gmail.com}
\emailAdd{probirrana@gmail.com}


\abstract{General lowest order perturbations to hermitian
squared mass matrices of leptons are considered away from the
tribimaximal (TBM) limit in which a weak flavor basis with mass
diagonal charged leptons is chosen. The three measurable TBM deviants
are expressed linearly in terms of perturbation  induced
dimensionless coefficients appearing in the charged lepton and
neutrino flavor eigenstates. With unnatural cancellations assumed to be
absent and the charged lepton perturbation contributions to their
flavor eigenstates argued to be small, we analytically derive the following
result. Within lowest order perturbations, a
deviation from maximal atmospheric neutrino mixing and the amount of 
CP violation in neutrino oscillations cannot both be large (i.e. $12$-$17 \% $), 
posing the challenge of verification to forthcoming experiments at the 
intensity frontier.

}

\maketitle
\flushbottom

\section{Introduction}

The phenomenon of mixing between different generations of quarks/leptons
has now been experimentally studied fairly well \cite{nf-1}. The three
quark mixing angles are at present quite well-measured. Though the leptonic
mixing angles are not known as well, one has credible nonzero $3 \sigma$
upper and lower bounds on them. CP violation has been investigated quite
thoroughly in the quark sector, but as yet there is no
reliable observation of CP-violation involving only leptons.
Quark mixing angles are known to become progressively smaller
in order of magnitude as one moves from $1$-$2$ to $2$-$3$ and
$1$-$3$ generation mixing. This fact can be understood qualitatively
in terms of a hierarchical quark mass matrix. The mixing angles, that
emerge from such a mass matrix, are small and turn out to be
given roughly by the mass ratios of relevant generations
of quarks. Since the masses of both up- and down-type quarks
are strongly hierarchical with respect to generations, this ties in with
observation. In complete contrast, the leptonic mixing angles have
been found to be much larger and show a different pattern. The
qualitative difference between quark and lepton mixing patterns is made
starkly evident by a quantitative comparison of the
approximate magnitudes \cite{nf-2,nf-3a,nf-3b,nf-3c} of the elements
of the respective unitary matrices $V_{CKM}$ and $U_{PMNS}$:
\begin{equation}
|V_{CKM}| \sim \begin{pmatrix}0.9 & 0.2 & 0.004 \cr
                      0.2 & 0.9 & 0.01 \cr
                      0.008 & 0.04 & 0.9 \end{pmatrix},
|U_{PMNS}| \sim \begin{pmatrix}0.8 & 0.5 & 0.2 \cr
                      0.4 & 0.6 & 0.7 \cr
                      0.4 & 0.6 & 0.7 \end{pmatrix}. \label{eqn-1}
\end{equation}

Though the masses of the charged leptons $l~(=e,\mu,\tau)$ show a
pronounced hierarchical pattern with respect to generations, one suspects
that such may not be the case with neutrinos. What operates for the
mixing the latter, possibly related to their presumed Majorana nature\footnote{We follow 
the procedure of Ref. \cite{nf-2} and take neutrinos
to be light Majorana particles occurring in three generations. Consequently, we
take a complex symmetric mass matrix for them. In the mass basis, that is
$M_\nu={\rm diag}.~(m_{\nu 1}, m_{\nu 2}, m_{\nu 3})$ with $m_{\nu 1}=|m_{\nu 1}|,~ 
m_{\nu 2}= |m_{\nu 2}| e^{- i \alpha_{21}},~ m_{\nu 3}=|m_{\nu 3}| e^{-i \alpha_{31}}$ and
$\alpha_{21},~\alpha_{31}$ as Majorana phases. We also use $c_{ij} \equiv \cos \theta_{ij}$ and
$s_{ij} \equiv \sin \theta_{ij}$ for the angle of mixing $\theta_{ij}$ between
neutrino flavors $i$ and $j$.} originating, say from some kind of a 
seesaw mechanism \cite{nf-2}, 
is perhaps some 
underlying family symmetry. Though one need not make
any specific assumption on the neutrino mass hierarchy, such considerations
are most natural for quasi-degenerate neutrinos. Even if there is
any mass hierarchy among neutrinos, it can be presumed to be
quite mild. Thus we separate the issue of the mixing of neutrinos from that of their mass 
hierarchy. More definitely, the
family symmetry controlling their mixing can be taken to be independent of the neutrino
mass hierarchy.  

For fermions of type $t~(=u,d,l,\nu)$, we can define the mass basis as
 one in which the corresponding mass matrix $M_t$ is diagonal.
We can also consider the flavor basis in which the fermions $|\chi^t \rangle$ are
flavor eigenstates but the mass matrix $M_{t f}$ is not necessarily 
diagonal. The hermitian squared mass matrix $M^\dagger M$ in each basis 
is related by a unitary transformation $U_t$:
\begin{equation}
U^\dagger_t M^\dagger_{t f} M_{t f} U_t=M^\dagger_t M_t. \label{eqn-2}
\end{equation}
We subscribe to the following viewpoint. While each of $U_u,U_d,U_\ell$ shows
a hierarchical structure, this is not true of $U_\nu$ which is
governed by a different principle. The way to gain new insights into
this principle is through more precise measurements of the leptonic
mixing angles and of the associated CP-violating Dirac phase $\delta_{CP}$ as
well as of the concerned neutrino masses. These can test mixing constraints
from specific theoretical ideas. Our aim in this paper is to
derive some such constraint which is experimentally testable. This we
do by considering lowest order perturbation theory in the additive
breaking of tribimaximal (TBM) neutrino mixing for  neutrino and
charge lepton mass matrices in the flavor basis. The additively broken TBM
paradigm is explained in detail below. From our consideration, we
obtain two alternative experimentally testable possibilities, at least one
of which is obligatory. Though our result is derived by use of general
arguments, we check it in specific flavor models.

The rest of the paper is organized as follows. Section 2 is devoted
to a discussion of tribimaximal mixing and its breaking. In section 3 we
set up our basic lowest order perturbative formalism which
is meant to compute the deviations away from 
tribimaximality. Section 4 contains the derivation of the
theoretical consequences of the said formalism. In Section 5 we discuss
the experimentally testable constraint arising therefrom. 
Section 6 includes a comparative study of our result with those of
various flavor models incorporating deviations from TBM.
The final section 7 summarizes our conclusions.

\section{Broken tribimaximal mixing and its effects}

There is a vast literature~\cite{nf-2, nf-3a,nf-3b,nf-3c,nf-4} covering theoretical
ideas on the principle governing $M_{\nu f}$ and $U_\nu$. Our focus, however, is
on tribimaximal (TBM) mixing~\cite{nf-4,nf-5a,nf-5b,nf-5c} which is elegant, predictive
and can be given a solid theoretical foundation~\cite{nf-6,nf-7} from specific
realizations of discrete family symmetries such as $A_4,S_3$ and $\Delta_{27}$.
Some of the latter have also suggested a few neutrino 
mixing sum-rules~\cite{nf-8na,nf-8nb,nf-8a,nf-8b,nf-8c,nf-8nc}. We henceforth use the superscript zero to 
denote the TBM limit. In this limit we choose
to work in the weak flavor basis in which the charged leptons have a diagonal
Dirac mass matrix
\begin{equation}
M^0_\ell={\rm diag.}~(m^0_e, m^0_\mu, m^0_\tau). \label{eqn-3}
\end{equation}
The TBM limit of the neutrino mass matrix $M^0_{\nu f}$ in the flavor
basis is characterized by certain linear relations among elements of
$M^0_{\nu f}$:
\begin{eqnarray}
(M^0_{\nu f})_{12} &=& -(M^0_{\nu f})_{13}, \label{eqn-4}\\
(M^0_{\nu f})_{22} &=& (M^0_{\nu f})_{33}, \label{eqn-5}\\
(M^0_{\nu f})_{11}-(M^0_{\nu f})_{13} &=& 
(M^0_{\nu f})_{22}-(M^0_{\nu f})_{23}. \label{eqn-6}
\end{eqnarray}
Given (\ref{eqn-4}),~(\ref{eqn-5}) and (\ref{eqn-6}), the most general
form of $M^0_{\nu f}$ is
\begin{equation}
M^0_{\nu f} = \begin{pmatrix}X & Y & -Y \cr
                 Y & X+Z & -Y+Z \cr
                 -Y & -Y+Z & X+Z \end{pmatrix}, \label{eqn-7}
\end{equation}
where $X,Y,Z$ are unknown complex mass dimensional parameters. Now the TBM pattern
obtains with the three emergent pairwise mixing angles, that appear \cite{nf-1} in $U_{PMNS}$, being
fixed at $
\theta^0_{12}=\sin^{-1} \sqrt{1 \over 3} \simeq 35.3^\circ,
\theta^0_{23}=\sin^{-1} \sqrt{1 \over 2} = 45^\circ,
\theta^0_{13}=0
$ independent of whether the neutrino mass ordering is normal or inverted.

We can compare the TBM-predicted values of the three mixing angles with their
current $3 \sigma$ allowed ranges. Recent global
fits yield \cite{nf-9,nf-10,nf-11} $31^\circ \le \theta_{12} \le 36^\circ,
36^\circ \le \theta_{23} \le 55^\circ, 7.2^\circ \le \theta_{13} \le 10^\circ$. 
Thus while $\theta_{12}$ and $\theta_{23}$ are certainly 
compatible with TBM values within their
measured ranges, $\theta_{13}$ a fortiori is not. Indeed, the measurement
of a significantly nonzero value of $\theta_{13}$ has been a major
experimental advance recently \cite{nf-12a,nf-12b,nf-12c,nf-12d} with a tremendous
theoretical impact. This is due to two reasons. First, CP-violation, that is observable
in neutrino oscillations, enters through the 
terms $s_{13}~e^{\pm i \delta_{CP}}$; thus $s_{13} \sim 0.12$-$0.17$ is
very encouraging to that end. Second, it means that any symmetry, leading to TBM, must
be a broken symmetry. The next natural question is: how quantitative is this breaking
and is TBM still relevant in an approximate sense ?

We make an attempt to answer this last question. Our approach is to add small
general perturbations to the TBM limits of hermitian
squared mass matrices $M^\dagger_{\ell f} M_{\ell f}$ and 
$M^\dagger_{\nu f} M_{\nu f}$. We take both sets of perturbations
to be of the same order of magnitude and treat them to the 
lowest order. Much effort~\cite{nf-13new,nf-13a,nf-13b,nf-13c} has already
been expended in this direction. However, we do have something new and 
interesting to say. We bring out
a novel feature of the near-TBM mixing of neutrinos
in terms of an analytically derived constraint which merits
being highlighted. The constraint implies that at least one of 
two conditions, that are testable in forthcoming neutrino oscillation 
experiments, must hold. Either the deviation $|s_{23}-\sqrt{1 \over 2}|$
from the maximal value of $\theta_{23}$  or the measure of CP violation 
$|s_{13} \sin \delta_{CP}|$ has to be quite  small ( $< 3 \%$ as opposed to\footnote{
We shall throughout refer to a TBM deviating effect as (1) ``very large'' if it
is $>>$ $100 s_{13} \% \sim $12-17$\%$ so that 
higher order perturbations cannot be ignored, (2) ``large'' if
it is in the ballpark of $100 s_{13}\%$ 
so that it should be soon measurable as well as computable with only
lowest order perturbations 
and (3) negligibly ``small''
if it is $< 0.03$ which is $O(s^2_{13})$.} $12$-$17 \%$ for the value of $s_{13}$ as compared with
unity),  
the latter meaning that CP will be conserved at the lowest order. This
conclusion is a consequence of the fact that the perturbed
eigenstates $|\chi^{\ell,\nu}_i \rangle$ for $i=1,2,3$ make up
the columns of the matrices $U_{\ell,\nu}$ to the lowest order.
Hence any observation in the near future of both a sizable deviation from 
maximal atmospheric 
neutrino mixing and a large amount of CP-violation
in neutrino oscillations would go against the
idea of lowest order additive perturbation to TBM-invariant
neutrino and charged lepton mass matrices.

In deriving the above conclusion, we do not assume any additional model 
either at a high or at a low
scale, or any specific discrete family symmetry. In fact, we perform
a lowest order model independent analysis with the most general
TBM violating perturbation matrices whose nonzero elements are
expected to be of the same order of magnitude. Moreover, our results on
neutrino mixing do not need to assume anything about the neutrino mass
ordering. This is since the perturbations are expected to be some kind of symmetry
breaking terms, which characterize their contributions to
$|\chi^{\ell,\nu} \rangle$ by a set of small dimensionless coefficients
$\{\epsilon^{\ell,\nu}\}$. All members of the subset $\{\epsilon^\nu\}$ in
the neutrino sector are taken to be typically of 
magnitude $\sim s_{13} \equiv \sin \theta_{13} \sim 0.12$-$0.17$, i.e. of the
order of $12$ -$17 \%$ or thereabouts of the unperturbed quantities. On the
other hand, in the charged lepton sector, arguements are given why $\{\epsilon^\ell\}$ are 
much less in magnitude than $\{\epsilon^\nu\}$ on account
of the strongly mass hierarchical nature of the charged leptons. 
This will be shown to follow from all nonzero perturbation matrix elements 
being taken to be of the same order of magnitude.
Of course, the neglected $O(\epsilon^2)$ terms 
are estimated to be
only at a $2$-$3 \%$ level which is below \cite{nf-new-1,nf-new-2} the accuracy of 
the measurement of TBM deviants in ongoing and forthcoming neutrino oscillation 
experiments\footnote{Experiments in the far future with neutrino factories
may probe such a level and, for such measurements, the neglected $O(\epsilon^2)$
effects as well as those due to renornalization group evolution from an
assumed high scale symmetry would be relevant.}. 

\section{Lowest order perturbation away from tribimaximality}

For charged leptons $\ell$ the normalized eigenvectors in the mass basis
and the flavor basis are identical in the TBM limit. Thus we can take
\begin{equation}
|\chi^{\ell 0}_1 \rangle=|\chi^{\ell 0}_1 \rangle_f=\begin{pmatrix}
1 \cr 0 \cr 0 \end{pmatrix},
|\chi^{\ell 0}_2 \rangle=|\chi^{\ell 0}_2 \rangle_f=
\begin{pmatrix}0 \cr 1 \cr 0 \end{pmatrix},
|\chi^{\ell 0}_3 \rangle=|\chi^{\ell 0}_3 \rangle_f=
\begin{pmatrix}0 \cr 0 \cr 1 \end{pmatrix}. \label{eqn-8}
\end{equation}
Moreover, the charged lepton mass matrix is identical in each basis in the
same limit, namely
\begin{equation}
M^0_{\ell f}=M^0_{\ell}. \label{eqn-9}
\end{equation}
Adding a perturbation $M^\prime_{\ell f}(\equiv \lambda_{ij})$ to $M^0_{\ell f}$ so that
$M_{\ell f}=M^0_{\ell f}+ M^\prime_{\ell f}$, we can construct
the corresponding matrix $M^\prime_\ell$ in the mass basis as 
\begin{equation}
M_\ell=M^0_\ell+M^\prime_\ell \label{eqn-10}
\end{equation}
via,
\begin{equation}
M^\dagger_\ell M_\ell=U^\dagger_\ell M^\dagger_{\ell f} M_{\ell f} U_\ell. \label{eqn-11}
\end{equation}

Turning to neutrinos in the TBM limit, we can write
\begin{equation}
{U^0_\nu}^\dagger {M^0_{\nu f}}^\dagger M^0_{\nu f} U^0_\nu
={\rm diag}.~(|m^0_{\nu 1}|^2,|m^0_{\nu 2}|^2,|m^0_{\nu 3}|^2) \label{eqn-12}
\end{equation}
with
\begin{equation}
U^0_\nu= \begin{pmatrix} \sqrt{2/3} & \sqrt{1/3} & 0 \cr
       -\sqrt{1/6} & \sqrt{1/3} & \sqrt{1/2} \cr
       \sqrt{1/6} & -\sqrt{1/3} & \sqrt{1/2} \end{pmatrix}. \label{eqn-13}
\end{equation}
The normalized flavor eigenvectors $|\chi^{\nu 0}_i \rangle$ of 
${M^0}^\dagger_{\nu f}M^0_{\nu f}$ for $i=1,2,3$ are the columns of
$U^0_\nu$ while those in the mass basis are identical to the
charged lepton ones. Thus
\begin{equation}
|\chi^{\nu 0}_1 \rangle=\begin{pmatrix}1 \cr 0 \cr 0 \end{pmatrix},
|\chi^{\nu 0}_2 \rangle=\begin{pmatrix} 0 \cr 1 \cr 0 \end{pmatrix},
|\chi^{\nu 0}_3 \rangle=\begin{pmatrix}0 \cr 0 \cr 1 \end{pmatrix}, \label{eqn-14}
\end{equation}
whereas
\begin{equation}
|\chi^{\nu 0}_1 \rangle_f=\begin{pmatrix}\sqrt{2 \over 3} \cr -\sqrt{1 \over 6} \cr \sqrt{1 \over 6} \end{pmatrix},
|\chi^{\nu 0}_2 \rangle_f=\begin{pmatrix}\sqrt{1 \over 3} \cr \sqrt{1 \over 3} \cr -\sqrt{1 \over 3} \end{pmatrix},
|\chi^{\nu 0}_3 \rangle_f=\begin{pmatrix}0 \cr \sqrt{1 \over 2} \cr \sqrt{1 \over 2} \end{pmatrix}. \label{eqn-15}
\end{equation}
Once the perturbation is introduced, we have $M_{\nu f}=M^0_{\nu f} + M^\prime_{\nu f}$,
where $M^0_{\nu f}$ obey the TBM conditions (\ref{eqn-4})--(\ref{eqn-6}) while 
$(M^\prime_{\nu f})_{ij} \equiv \mu_{ij}=\mu_{ji}$ violate them. The violation in TBM conditions is
given by,
\begin{eqnarray}
(M_{\nu f})_{12}+(M_{\nu f})_{13} &=& \mu_{12}+\mu_{13}, \label{new1}\\
(M_{\nu f})_{22}-(M_{\nu f})_{33} &=& \mu_{22}-\mu_{33}, \label{new2}\\
(M_{\nu f})_{11}-(M_{\nu f})_{13}-(M_{\nu f})_{22}+(M_{\nu f})_{23} &=& \mu_{11}-\mu_{13}-\mu_{22}+\mu_{23}. \label{new3}
\end{eqnarray}
Note that, unlike the real diagonal
$M^0_\ell$ and the general $M^\prime_\ell$, both $M^0_{\nu f}$ and $M^\prime_{\nu f}$ 
have to be complex symmetric matrices in order to make the corresponding neutrinos 
Majorana particles.

We now expand the perturbed eigenstates for both charged
leptons and neutrinos at the lowest order. We choose to use a compact
notation covering both  cases by introducing perturbation parameters
$\epsilon^{\nu,\ell}_{ik} $ (for $i,k=1,2,3$). Thus we can write the ith first order perturbed
eigenvectors of $M^\dagger_{\nu f}M_{\nu f}$ 
on one hand and of $M^\dagger_{\ell f}M_{\ell f}$ on the other as  
\begin{equation}
|\chi^{\nu,\ell}_{i}\rangle_{f} = |\chi^{0 \nu,\ell}_{i}\rangle_{f} 
+\sum_{k \ne i}\epsilon^{\nu,\ell}_{ik} | \chi^{0 \nu,\ell}_{k} \rangle_{f} 
+ O(\epsilon^2).\label{eqn-16}
\end{equation}
Two new quantities have been introduced in (\ref{eqn-16}). They are defined by
\begin{eqnarray}
\epsilon^{\nu,\ell}_{ik} 
&=& -{\epsilon^{\nu,\ell}}^*_{ki}=
(|m^0_{\nu,\ell i}|^2-|m^0_{\nu,\ell k}|^2)^{-1}p^{\nu,\ell}_{ki},\label{eqn-17}\\
p^{\nu,\ell}_{ik} &=& \langle \chi^{0 \nu,\ell}_i|{M^0_{\nu,\ell}}^\dagger M^\prime_{\nu,\ell} 
+ {M^\prime_{\nu,\ell}}^\dagger M^0_{\nu,\ell}|\chi^{0 \nu,\ell}_k \rangle. \label{eqn-18}
\end{eqnarray} 
Note that (\ref{eqn-17}) and (\ref{eqn-18}) have been written in the mass basis
utilizing the fact that $\epsilon^{\nu,\ell}_{ik}$, as well as $p^{\nu,\ell}_{ik}$, do not 
change from one basis to the other. We can also comment on the lack of dependence of the epsilon
 parameters on the yet unknown overall
neutrino mass scale. If $M^0_\nu$ and $M^\prime_\nu$ are both scaled by a 
factor $\alpha$, the unperturbed eigenvalues $\{ m^0_{\nu i} \}$ will also be scaled
similarly. As a result, $\epsilon^\nu_{ik}$ will remain invariant under an overall mass 
scaling. On the other hand, suppose two of the mass eigenvalues are large but close to one another, as is
the case with $\nu_1$ and $\nu_2$ with an inverted mass hierarchy, and this is not much affected
by the perturbations. In such a case the corresponding $\epsilon^\nu_{12}$ will get enhanced.

Turning to (\ref{eqn-16}), we see that its LHS for $i=1,2,3$ can be identified
with three corresponding columns of $U_{\nu,\ell}$. Thus
\begin{equation}
U_{\nu,l}=(|\chi^{\nu,\ell}_1 \rangle_f ~~|\chi^{\nu,\ell}_2 \rangle_f ~~|\chi^{\nu,\ell}_3 \rangle_f). 
\label{eqn-19}
\end{equation}
Neglecting $O(\epsilon^2)$ terms, it follows from (\ref{eqn-19}) that
\begin{equation}
U_\ell=\begin{pmatrix}1 & -\epsilon^{\ell *}_{12} & -\epsilon^{\ell *}_{13} \cr
                              \epsilon^{\ell }_{12} & 1 & -\epsilon^{\ell *}_{23} \cr
                              \epsilon^{\ell }_{13} & \epsilon^{\ell }_{23} & 1 \end{pmatrix}
\label{eqn-20}
\end{equation}
and
\begin{equation}
U_\nu= 
\begin{pmatrix}
\sqrt{\frac{2}{3}}+ \sqrt{\frac{1}{3}} \epsilon^\nu_{12} 
& \sqrt{\frac{1}{3}}-\sqrt{\frac{2}{3}} \epsilon^{\nu *}_{12} 
& -\sqrt{\frac{2}{3}}\epsilon^{\nu *}_{13} -\sqrt{\frac{1}{3}} \epsilon^{\nu *}_{23} \cr
-\sqrt{\frac{1}{6}}+ \sqrt{\frac{1}{3}} \epsilon^\nu_{12}+\sqrt{\frac{1}{2}} \epsilon^\nu_{13} 
& \sqrt{\frac{1}{3}}+\sqrt{\frac{1}{6}} \epsilon^{\nu *}_{12}+\sqrt{\frac{1}{2}} \epsilon^\nu_{23} 
& \sqrt{\frac{1}{2}} + \sqrt{\frac{1}{6}}\epsilon^{\nu *}_{13} -\sqrt{\frac{1}{3}} \epsilon^{\nu *}_{23} \cr
\sqrt{\frac{1}{6}}- \sqrt{\frac{1}{3}} \epsilon^\nu_{12}+\sqrt{\frac{1}{2}} \epsilon^\nu_{13} 
& -\sqrt{\frac{1}{3}}-\sqrt{\frac{1}{6}} \epsilon^{\nu *}_{12}+\sqrt{\frac{1}{2}} \epsilon^\nu_{23} 
& \sqrt{\frac{1}{2}}-\sqrt{\frac{1}{6}} \epsilon^{\nu *}_{13} + \sqrt{\frac{1}{3}} \epsilon^{\nu *}_{23}
\end{pmatrix}.
\label{eqn-21}
\end{equation}

Let us define the Majorana phase matrix
\begin{equation}
K \equiv {\rm diag}.~(1,e^{i \alpha_{21} \over 2},e^{i \alpha_{31} \over 2}).
\label{eqn-22}
\end{equation}
Then $U_{PMNS}K^{-1}$ can be written in the PDG convention \cite{nf-1} as
\begin{equation}
U_{PMNS}K^{-1} =
\begin{pmatrix}
c_{12}c_{13} & s_{12}c_{13} & s_{13}e^{-i \delta_{CP}} \cr
-s_{12}c_{23}-c_{12}s_{23}s_{13}e^{i \delta_{CP}} & c_{12}c_{23}-s_{12}s_{23}s_{13}e^{i \delta_{CP}} & s_{23} c_{13} \cr
s_{12} s_{23} -c_{12} c_{23} s_{13} e^{i \delta_{CP}} & -c_{12}s_{23}-s_{12}c_{23}s_{13} e^{i \delta_{CP}} & c_{23}c_{13}
\end{pmatrix}. \label{eqn-23}
\end{equation}
We can now make the identification
\begin{equation}
U_{PMNS}K^{-1}=U^\dagger_\ell U_\nu \label{eqn-24}
\end{equation}
and work out the consequences from (\ref{eqn-20}), (\ref{eqn-21}) and (\ref{eqn-23}).

\section{Consequences of lowest order perturbation}
Let us define $L \equiv U^\dagger_\ell U_\nu$ and $N \equiv U_{PMNS} K^{-1}$. The identification
$L_{ij}=N_{ij}$ as per (\ref{eqn-24}) leads to nine equations which are detailed in
convenient combinations in the Appendix. Not all of these are independent, but they
lead to four independent constraint conditions and three equations for the three
TBM-deviants $c_{12}-\sqrt{2 \over 3}, c_{23}-s_{23}$ and $s_{13} e^{i \delta_{CP}}$. The constraint
conditions follow from the fact that four of the elements of $N$ are real. They are
given by
\begin{eqnarray}
&& {\rm Im}~\epsilon^\nu_{12} = O(\epsilon^2), \label{eqn-25} \\
&& {\rm Im}~(\epsilon^\nu_{13}-\sqrt{2} \epsilon^\nu_{23}) = O(\epsilon^2), \label{eqn-26}\\
&& {\rm Im}~\epsilon^l_{23} = O(\epsilon^2) \label{eqn-27}, \\
&& {\rm Im}~(\epsilon^l_{12}-\epsilon^l_{13}) = O(\epsilon^2). \label{eqn-28}
\end{eqnarray}
Neglecting $O(\epsilon^2)$ terms, the three measurable TBM-deviants are linear in the
$\epsilon$ coefficients and may be given as
\begin{eqnarray}
&& c_{12}-\sqrt{2 / 3} 
 =\sqrt{1 / 2}\left(\sqrt{1 / 3} -s_{12} \right)
 =\sqrt{1 / 3} ~\epsilon^\nu_{12} 
- \sqrt{1 / 6}\left( \epsilon^l_{12}-\epsilon^l_{13} \right),\label{eqn-29} \\
&& c_{23}-s_{23} 
 =-\sqrt{2 / 3}\left(\epsilon^\nu_{13}-\sqrt{2} 
~\epsilon^\nu_{23}\right)-\sqrt{2}~ \epsilon^l_{23},  \label{eqn-30} \\
&& s_{13}~e^{i \delta_{CP}} 
 =-\sqrt{1 / 3}~\left(\sqrt{2}~ \epsilon^\nu_{13} 
+ \epsilon^\nu_{23}\right)
+\sqrt{1 / 2}\left(\epsilon^l_{12}+\epsilon^l_{13}\right). \label{eqn-31}
\end{eqnarray}
The derivation of Eqs. (\ref{eqn-25})--(\ref{eqn-31}) appears in the Appendix.

Because of (\ref{eqn-26}), the real and imaginary parts of (\ref{eqn-31}) enable
us to write, modulo $O(\epsilon^2)$ terms, that
\begin{equation}
\tan \delta_{CP}
= \frac{3 ~{\rm Im}~\epsilon^\nu_{23} -\sqrt{3 /2} ~{\rm Im}~(\epsilon^\ell_{12}+\epsilon^\ell_{13})
}{
{\rm Re}~(\sqrt{2} ~\epsilon^\nu_{13} + \epsilon^\nu_{23})
-\sqrt{3 / 2} ~{\rm Re}~(\epsilon^l_{12} + \epsilon^l_{13})}. \label{eqn-32}
\end{equation}
The above equation may be recast in terms of the basis independent Jarlskog 
invariant $J$ which equals 
\[ 
{\rm Im}~[(U_\ell^\dagger U_\nu)_{e 1}(U_\ell^\dagger U_\nu)_{\mu 2}
(U_\ell^\dagger U_\nu)^*_{e 2}(U_\ell^\dagger U_\nu)^*_{\mu 1}].
\]
We then have
\begin{equation}
J=-{\frac{1}{\sqrt{6}}}~{\rm Im}~[\epsilon^\nu_{23}-{\frac{1}{\sqrt{6}}}
(\epsilon^l_{12} + \epsilon^l_{13})]+O(\epsilon^2). \label{eqn-33}
\end{equation}

Let us now explore, to the lowest order in $\epsilon$, the consequences of
(\ref{eqn-17}) and (\ref{eqn-18}) by explicitly taking elements of the
respective perturbing mass matrices for neutrinos and charged leptons. We take
\begin{equation}
(M^\prime_{\nu f})_{ij}= \mu_{ij}=\mu_{ji} \label{eqn-34}
\end{equation}
 and 
\begin{equation}
(M^\prime_{\ell f})_{ij}=(M^\prime_\ell)_{ij}+O(\epsilon^2) \label{eqn-35} 
=\lambda_{ij}
\end{equation}
with $\lambda_{ij}$ and $\mu_{ij}=\mu_{ji}$ as complex mass dimensional 
parameters naturally expected to be of the same order of magnitude.  The identity of the charged lepton 
mass basis and flavor basis
in the TBM limit makes the calculations in this case quite
straightforward. From (\ref{eqn-17}) and (\ref{eqn-18}), we can easily derive
\begin{eqnarray}
\epsilon^l_{12} &=& ({m^0_e}^2-{m^0_\mu}^2)^{-1}~(m^0_\mu \lambda_{21} + m^0_e \lambda^*_{12}), \label{eqn-36}\\
\epsilon^l_{23} &=& ({m^0_\mu}^2-{m^0_\tau}^2)^{-1}~(m^0_\tau \lambda_{32} + m^0_\mu \lambda^*_{23}), \label{eqn-37}\\
\epsilon^l_{13} &=& ({m^0_e}^2-{m^0_\tau}^2)^{-1}~(m^0_\tau ~\lambda_{31}+m^0_e{\lambda^*_{13}}). \label{eqn-38}
\end{eqnarray}
We want to comment on the magnitudes of $\epsilon^\ell_{23}$ and $\epsilon^\ell_{13}$.
In order for them to be large, the relevant $\lambda$ parameters
would need to be of order $m_\tau$. That is not in conformity with our premise
that nonzero charged lepton perturbation mass matrix elements (
i.e. $\lambda_{ij}$) 
cannot be very different in order
of magnitude from those for neutrinos (i.e. $\mu_{ij}$). Thus we expect that $|\epsilon^\ell_{23}|$ 
and $|\epsilon^\ell_{13}|$ to be
quite small. 
In any event, because of the strongly hierarchical nature of charged 
lepton masses, (\ref{eqn-27}) and
(\ref{eqn-28}) can be satisfied without unnatural cancellations
only by $\lambda_{12},\lambda_{21},\lambda_{13},\lambda_{31},\lambda_{23},\lambda_{32}$
all being real to order $\epsilon$. One then automatically obtains that
\begin{equation}
{\rm Im} ~\epsilon^l_{12}=O(\epsilon^2)={\rm Im}~ \epsilon^l_{13}.\label{eqn-39}
\end{equation}
Feeding this information, we can simplify (\ref{eqn-32}) and (\ref{eqn-33}) to
\begin{equation}
\tan \delta_{CP}
=\frac{3 ~{\rm Im}~\epsilon^\nu_{23} 
}{
{\rm Re}~(\sqrt{2} ~\epsilon^\nu_{13} + \epsilon^\nu_{23})
-\sqrt{3 / 2} ~ {\rm Re}~(\epsilon^l_{12} + \epsilon^l_{13})},
\label{eqn-40}
\end{equation}
\begin{equation}
J=-\frac{1}{\sqrt{6}}~{\rm Im}~\epsilon^\nu_{23} +O(\epsilon^2)\label{eqn-41}
\end{equation}
respectively.

Turning to  neutrinos next, the relevant off-diagonal
elements of $M^\prime_\nu={U^0_\nu}^T M^\prime_{\nu f} U^0_\nu$ are
\begin{eqnarray}
(M^\prime_\nu)_{12} &=&
\frac{1}{3 \sqrt{2}}(2 \mu_{11}+\mu_{12}-\mu_{13}-\mu_{22}+2 \mu_{23}-\mu_{33}), \label{eqn-42}\\
(M^\prime_\nu)_{23}
&=& \frac{1}{\sqrt{6}}(\mu_{12}+ \mu_{13}+ \mu_{22} - \mu_{33}), \label{eqn-43}\\
(M^\prime_\nu)_{13}
&=& \frac{1}{\sqrt{3}}(\mu_{12}+ \mu_{13}-\frac{1}{2} \mu_{22} + \frac{1}{2}
\mu_{33}). \label{eqn-44}
\end{eqnarray}
It is now convenient to define
\begin{eqnarray}
&& \Delta^0_{ij} \equiv |m^0_{\nu i}|^2-|m^0_{\nu j}|^2, \label{eqn-45} \\
&& a^\mp_{ij} \equiv m^0_{\nu i} \mp m^0_{\nu j}. \label{eqn-46}
\end{eqnarray}
Then we take (\ref{eqn-17}) and (\ref{eqn-18}) and successively
consider the index combinations $i=1,~k=2$ and $i=2,~k=3$ as well as $i=1,~k=3$. Separating
the real and imaginary parts and using (\ref{eqn-45}) and (\ref{eqn-46}), we obtain
the following six equations
\begin{eqnarray}
 2~\Delta^0_{12}~
\begin{pmatrix} {\rm i ~Im}~\epsilon^\nu_{12} \cr
{\rm Re}~\epsilon^\nu_{12} \end{pmatrix} 
&&=
{a^\mp}^*_{21}~{(M^\prime_\nu)}_{12}
\mp c.c.,\label{eqn-47}\\
&& \nonumber\\
 2~\Delta^0_{23}~
\begin{pmatrix} {\rm i~Im}~\epsilon^\nu_{23} \cr
  {\rm Re}~ \epsilon^\nu_{23} \end{pmatrix}
&&= {a^\mp}^*_{32}{(M^\prime_\nu)}_{23}
 \mp c.c., \label{eqn-48} \\
&& \nonumber\\
 2 ~ \Delta^0_{13}~
\begin{pmatrix} {\rm i~Im}~\epsilon^\nu_{13} \cr
 {\rm Re}~\epsilon^\nu_{13} \end{pmatrix}
&&= {a^\mp}^*_{31}(M^\prime_\nu)_{13} \mp c.c. \label{eqn-49}
\end{eqnarray}
Needless to add, order $\epsilon^2$ terms have been neglected in deriving
the above results.

\section{Results and discussion}

Eq. (\ref{eqn-47}) has a simple consequence if we exclude
unnatural cancellations. In conjunction with (\ref{eqn-25}), it forces
the combination of $\mu_{ij}$, occurring in $(M^\prime_\nu)_{12}$, i.e. 
$2 \mu_{11}+ \mu_{12}-\mu_{13}-\mu_{22} + 2 \mu_{23}-\mu_{33}$, to be
real. It also implies that $m^0_{\nu_2}-m^0_{\nu_1}$ is real, the latter forcing
$\alpha^0_{21}$ to be $0$ or $\pi$. However, our key observation follows from
combining (\ref{eqn-48}) and (\ref{eqn-49}) with (\ref{eqn-26}). That
procedure yields the equality
\begin{eqnarray}
&& {\rm Im}~[ ({m^{0 *}_{\nu_3}}-{m^{0 *}_{\nu_2}})(\mu_{12}+\mu_{13}+\mu_{22}-\mu_{33})] \nonumber\\
&& =
{\rm Im}~[ ({m^{0 *}_{\nu_3}}-{m^{0 *}_{\nu_1}})(\mu_{12}+\mu_{13}+{1 \over 2}\mu_{22}-{1 \over 2}\mu_{33})].
\label{eqn-50}
\end{eqnarray}
There are two ways to satisfy (\ref{eqn-50}) without any unnatural
cancellation, at least one of which is obligatory.
Either we must have  option {\bf (1)}, namely that $m^0_{\nu_2}=m^0_{\nu_1}$ and $\mu_{22}=\mu_{33}$ 
or there must be option {\bf (2)}, namely that $m^0_{\nu_3},m^0_{\nu_2},m^0_{\nu_1},\mu_{12}+\mu_{13}$ and $\mu_{22}-\mu_{33}$
are all real so that each side of (\ref{eqn-50}) vanishes. Take {\bf (1)} first. Since $m^0_{\nu 1}=|m^0_{\nu 1}|$ by 
choice, we now have $|m^0_{\nu_1}|=|m^0_{\nu_2}|$ and $\alpha^0_{21}=0$,  
i.e. $\Delta_{21} \equiv |m_{\nu 2}^2|-|m_{\nu 1}|^2$ arises solely from TBM breaking. 
Further, with
$\mu_{22}=\mu_{33}$, the implication from from (\ref{eqn-48})
and (\ref{eqn-49}) is that $\sqrt{2}~{\rm Re}~\epsilon^\nu_{23}={\rm Re}~\epsilon^\nu_{13} 
+ O(\epsilon^2)$. Consequently, it follows  from (\ref{eqn-27}) and (\ref{eqn-30}) that
$c_{23}-s_{23}=-\sqrt{2} \epsilon^{\ell}_{23}+O(\epsilon^2)$ which leads to the result
$|s_{23}-{1 \over \sqrt{2}}|={1 \over \sqrt{2}} |\epsilon^\ell_{23}| 
+|O(\epsilon^2)| << |O(\epsilon^\nu)|$. The strong inequality in the last step 
has been based on the discussion which followed (\ref{eqn-38}).
Thus option {\bf(1)} says that the magnitude of any deviation from maximal atmospheric mixing, being of order
$|\epsilon^\ell_{23}|$ and small, will not be easily observed in forthcoming experiments. Let us turn to alternative {\bf (2)}. Now we have
$\alpha^0_{21}$ and $\alpha^0_{31}$ equalling $0$ or $\pi$. Further, by use of
(\ref{eqn-43}) and (\ref{eqn-48}), we derive that ${\rm Im}~\epsilon^\nu_{23}=O(\epsilon^2)$. As a result, by 
virtue of (\ref{eqn-40}) as well as (\ref{eqn-41}), one concludes that $s_{13}\sin \delta_{CP}=O(\epsilon^2)$
and $J=O(\epsilon^2)$, so that both would be small and hard to detect in
experiments planned for the near future. The implication  of option {\bf (2)} is that
CP violation in neutrino oscillations may not be seen 
in those experiments. It may be noted that the assumption $|\epsilon^\ell| << |\epsilon^\nu|$ is
unnecessary for this option.

It is also noteworthy that in option ({\bf 1}) one needs to use degenerate perturbation 
theory \cite{nf-14a,nf-14b,nf-14c} with respect to the TBM limit for the 1-2 sector 
of neutrinos. In the latter
case, the perturbation splits the 1-2 mass degeneracy and generates the solar
neutrino mass difference with $m^0_{\nu 1}=m^0_{\nu 2}=m^0_\nu$. One  then obtains
\begin{equation}
\Delta_{21}=\sqrt{(p^\nu_{11}-p^\nu_{22})^2+{p^\nu_{12}}^2},
\end{equation}
as calculated using (\ref{eqn-18}). Additionally, to order $\epsilon^\nu$ and $\epsilon^\ell$, 
$s_{13} e^{i \delta_{CP}}$ can be obtained
in terms of $m^0_{\nu 3},m^0_\nu,m^0_e,m^0_\mu,m^0_\tau$ as well as  the $\mu$ 
and $\lambda$ parameters
by using (\ref{eqn-31}) and employing the expressions for the $\epsilon$ parameters.
We choose not write that full expression here.

Some comments on the issue of unnatural cancellations are in order.
The TBM breaking terms in the mass matrix of charged leptons do not leave
any residual symmetry except possibly some rephasing invariances. As stated
earlier, given that $m_e << m_\mu << m_\tau$, the cancellations required to
avoid the reality condition on all $\lambda_{ij}$ (for $i \ne j$) cannot be
effected by any such invariance. In the neutrino case, there generally is a
residual $Z_2$ symmetry \cite{nf-5a,nf-5b,nf-5c,nf-6,nf-7,nf-8a,nf-8b}  
after TBM is broken. Even such a 
discrete symmetry does not generally enable one to obtain the 
concerned complicated
equality between specific combinations of TBM violating perturbation
parameters, TBM invariant neutrino masses as well as Majorana phases.
We feel, therefore, that our argument ruling out such cancellations
is sound and our conclusions are reliable.

 Let us finally remark on the relevance of our result to
planned experiments at the proton beam intensity frontier.
The determination of the sign of the neutrino mass ordering is one of
their aims. It is noteworthy that the constraint on neutrino mixing
parameters, derived 
by us, is independent of this issue 
{\it just as the consequences of exact TBM are}. Those 
experiments will also investigate
neutrino mixing parameters.  A combination \cite{nf-15a,nf-15b, nf-15c} of data from 
the ongoing and upcoming runs of 
T2K and NO$\nu$A experiments would probe 
$|s_{23}-{1 \over \sqrt{2}}|$ from the conversion
probability $P(\nu_\mu \rightarrow \nu_e)$. 
Now, in case a sizable nonzero value of that quantity is measured, 
being of magnitude comparable in percentage terms to (100 $s_{13}$)$\%$
of the maximal value of $s_{23}$, our condition ({\bf 2}) would hold and predict
a small amount  of CP violation in neutrino
oscillations from the above data. Contrariwise, the
failure to measure any  deviation  from maximal atmospheric neutrino mixing outside error bars
would mean that our condition ({\bf 1}) would operate with
$s_{13}\sin \delta_{CP}  = O(\epsilon^\nu)$, $J=O(\epsilon^\nu)$ permitted; that
would bolster the hope of detecting CP nonconservation for
oscillating neutrinos from the difference in conversion
probabilities $P(\nu_\mu \rightarrow \nu_e)
-P(\overline{\nu_\mu} \rightarrow \overline{\nu_e})$. The latter
would be good news not only for a combined analysis of
data from forthcoming runs of \cite{nf-16} of T2K and
NO$\nu$A but also for future experiments with
superbeams, such as LBNF \cite{nf-17}, LBNO \cite{nf-18a,nf-18b} or a neutrino factory
at 10 GeV \cite{nf-19}. Current hints, either
for a non-maximal $\theta_{23}$ or a nonzero $\sin \delta_{CP}/J$,  
by no means constitute any robust evidence and an experimental resolution of these two
issues is urgently called for.

\section{Comparative studies with specific flavor models
of broken TBM}

In the present analysis we have used first order perturbation theory
to analytically establish relations between basis independent sets of
small coefficients  $\{\epsilon^\ell, \epsilon^\nu\}$ and TBM deviant 
measurables. In doing so, we have been able to establish the relations
given in Eq.(\ref{eqn-25}) to Eq.(\ref{eqn-28}). Physical observables
partaining to CP violation have also been related analytically to
these basis independent $\epsilon$ coefficients. For these relations
to remain valid, TBM symmetry should be broken weakly so that
one could jusify first order perturbation theory. If that
symmetry is broken strongly, in other words, if the Lagrangian
contains large terms violating TBM symmetry, then these relations would
fail to be true
\footnote{ TBM braking in general is naturally expected to be
under control for lowest order perturbation theory since
$s_{13}$ has been observed to be $<0.18$.}
. In that case direct numerical diagonalization
would need to be made. On the other hand, numerical diagonalization
cannot be done in a model independent way;  consequently, a case by case
study, depending on the model of TBM symmetry breaking, would be required.

Given our two  assumptions, namely (1) $|\epsilon^\ell| << |\epsilon^\nu|$
and (2) the absence of unnatural cancellations, it is desirable to
cross check our result with specific flavor symmetry models which break
tribimaximality by some amount. We consider below several such proposed models
in a representative but not comprehensive survey. Most (though not all) of these are 
variations of a basic family symmetry
model \cite{nf-39} utilizing the discrete group $A_4$ along with
gauge singlet Higgs fields called flavons which transform as specified
$A_4$ representations. Not every such model can be cast within the
framework of additive perturbations to $M^0_{\nu f}$ and $M^0_{\ell f}$.
Nonetheless, we deem it useful to make this comparison. In these models, if
some flavons develop VEVs aligned in appropriately chosen directions in the
corresponding $A_4$ representation space, TBM obtains in the neutrino 
sector with mass diagonal charged leptons. Certain higher mass dimensional 
terms are entered into the Lagrangian containing ratios of flavon VEVs divided 
by a much larger cut-off scale. If some slight misalignment is then introduced 
in these VEV directions, deviations result from exact TBM.

The first example of this type on our list is that of Ref.\cite{nf-40} which
utilizes two $A_4$ triplet and three $A_4$ singlet flavons. An 
analytical study of this model was made here while the
TBM deviants were investigated numerically. The magnitude of the
dominant TBM breaking parameters was 
restricted to small values 
 by taking $|U_{13}| < 0.2$. This
analysis took care to ensure unitary implementation of broken TBM symmetry, i.e. that 
perturbed eigenstates of 
type $t$ do make up the columns of $U_t$. A revealing facet of this 
model is that the misalignment induced coefficients in the
perturbed charged lepton eigenstates turn out to be significantly
less in magnitude than the corresponding ones for neutrinos. This
is since the latter get enhanced by mass ratio factors such as
$(m^0_{\nu 1} + m^0_{\nu 3})(m^0_{\nu 2}-m^0_{\nu 1})^{-1},
(m^0_{\nu 1} + m^0_{\nu 3})|m^0_{\nu 2}-m^0_{\nu 3}|^{-1}$ and
$(m^0_{\nu 3} + m^0_{\nu 1})|m^0_{\nu 1}-m^0_{\nu 3}|^{-1}$ in our
notation, from the imposed unitary implementation. 
{\it The corresponding factors in the charged lepton case
are non-enhancing because of the hierarchical nature of the charged lepton masses.} 
Thus the model
manifestly satisfies our condition $|\epsilon^\ell| < |\epsilon^\nu|$.
The computed numerical values of $J$ are found to go all the
way up to $0.046$ when the full parameter space is scanned, cf.
Table II of Ref.\cite{nf-40}. This means 
that $(s_{13} \sin \delta_{CP})_{\rm max} \sim 0.195$, allowing
substantial possible CP violation in neutrino oscillations. However,
throughout the parameter space, one always has $\sin^2 2 \theta_{23} < 0.994$
i.e. $|s_{23} -\sqrt{1 \over 2}| < 0.03$, which permits only a tiny
deviation from maximality in atmospheric atmospheric neutrino mixig. Therefore this model satisfies our
option (1). The second model \cite{nf-41} that we consider is very
similar to that of Ref.\cite{nf-40} except that 
the perturbations can be arbitrarily large and  
real flavon VEVs were
chosen; consequently, there is no CP violation to be observed in neutrino
oscillations. The deviation from maximality in $s_{23}$ can be made large 
only by chosing the TBM breaking perturbation parameter $|U_{e3}| \sim 0.4$. If
$|U_{e3}|$ is restricted to $<0.2$, as dictated by later experiments,
once again the numerical constraint $|s_{23} - \sqrt{1 \over 2}| < 0.03$ is
seen to operate in agreement with Ref.\cite{nf-40}, i.e. the deviation
from maximal neutrino mixing is small by our critarion. Hence our option
(1) is maintained here with the additional proviso of a nonexistent $J$.
Significant deviations $|s_{23}-\sqrt{2 \over 3}|$ can occur for very large perturbations
which are beyond the scope of our work.

We then consider the study of $A_4$ and $S_4$ based flavor symmetry
models with perturbed lepton mass matrices reported in Ref.\cite{nf-42}. In
particular, for the $S_4$ based model investigated, $s^2_{23}$ gets fixed at $1/2$
and there is no deviation from maximality in atmospheric neutrino 
mixing; moreover, $\delta_{CP}$ is preferred to be near $\pi$, i.e. no
significant CP-violation in neutrino oscillations is predicted. So this
model is compatible with both our options $1$ and $2$. In the $A_4$ based
model considered (with just two $A_4$ singlet flavons), the authors derive the
sum rule $s^2_{13} \sin^2 \delta_{CP}=s^2_{13} -2 (s^2_{23}-1/2)^2$. It is
noteworthy that both  our options $1$ and $2$ are compatible with
this result. This is since, according to the sum rule, CP violation in
neutrino oscillations is largest when $s^2_{23}=1/2$ while the deviation
from maximal atmospheric neutrino mixing is greatest when the Dirac phase 
$\delta_{CP}=0$ or $\pi$ i.e. there is no CP violation.

The next analysis in our menu is that of Ref.\cite{nf-43}. Here again
$A_4$-based models are considered with the number of $A_4$ singlet
flavons varying from one to three and with the possibility of including
the see-saw mechanism for neutrino mass generation. Additive perturbations 
are considered $vis$-$a$-$vis$ TBM invariant charged
lepton and neutrino mass matrices and numerical diagonalization
is carried out.  The parameter spaces of the
models considered here allow both a substantial $J \sim 0.02$ 
(i.e. $\delta_{CP} \sim 30^\circ$) and a sizable $|s^2_{23} -1/2| \ge 0$.
However, unlike in Ref.\cite{nf-40}, very large perturbation
parameters have been alowed here. For instance, the charged
lepton perturbations $\epsilon^{ch}$ have been taken upto $0.3$
while the corresponding neutrino ones have been kept completely
free in the numerical scan with large allowed values. Thus lowest order perturbation 
theory does not apply to a
considerable region of their parameter space. We expect that their
results should agree with those of Ref.\cite{nf-40} once the smallness
criterion is imposed on the perturbations.

The final analysis within the ambit of our comparative study
is that of Ref.\cite{nf-44}. This work is somewhat different
from the previously considered models in that no specific flavor
symmetry such as $A_4$ for the Lagrangian is assumed. Instead, three
separate mechanisms of TBM breaking are 
considered $per~se:$ $(1)$ corrections
to $U_\ell$ in the charged lepton sector while 
keeping $U^0_\nu$ unchanged, $(2)$ renormalization group corrections
(with supersymmetry) starting from exact TBM and nearly mass degenerate
neutrinos at a very high scale and $(3)$ explicit TBM breaking
terms added to $M^0_{\nu f}$ in the neutrino sector only. For $(1)$, the
authors find that $J$ approaches a near maximum 
with $\delta_{CP} \sim \pi/2$ but the deviation from maximal atmospheric 
neutrino mixing is small with $s^2_{23}=1/2+ O(|U_{e3}|^2)$. This respects
our option $(1)$. For cases $(2)$ and $(3)$ of Ref.\cite{nf-44}, sizable
such deviations in the latter are possible with $|s_{23}^2-1/2| \sim 0.1$-$0.2$; 
however, $J$ was not investigated. For  case $(2)$, in particular, exact
TBM at a high scale makes the starting boundary value 
of $\delta_{CP}$ indeterminate and an unambiguous answer is not possible.

\section{Concluding summary}

In this paper we have considered general perturbations at the
lowest order to hermitian squared mass matrices $M^\dagger_{\ell f} M_{\ell f}$ and
$M^\dagger_{\nu f} M_{\nu f }$ respectively for charged leptons and neutrinos 
in the flavor basis of each and away from their
TBM limits by carefully taking into account the unitary relation between the
mass basis and the flavor basis. We have utilized the fact that columns of
the said unitary matrix are the perturbed eigenstates. We have derived
linear expressions for the three measurable TBM deviants in terms of the dimensionless coefficients
that appear in the perturbed charged lepton and neutrino eigenstates. 
We have further derived four independent constraints on the imaginary parts
of the latter from the requirement that four of the elements of $U^\dagger_\ell U_\nu$ have
to be real. With the plausible arguements of the mixing caused by the strongly mass hierarchical
charged leptons being significantly smaller than that due to neutrinos and
no unnatural cancellations, we have derived a result, forcing
one of two possibilities, which should be testable in the
foreseeable future. This main result of ours can be stated succintly
in the language of mathematical logic. Proposition A: an accurate
description of neutrino mixing is given by
the lowest order of additively perturbed
tribimaximality without unnatural cancellations and with the mixing from
the strongly mass hierarchical charged leptons being significantly smaller than that from neutrinos.
Proposition B: $|s_{23}-\sqrt{1 \over 2}|=O(\epsilon^\nu)$. Proposition C: 
$s_{13} \sin \delta_{CP}/J=O(\epsilon^\nu)$. Then $A \cap ( B \cup C) = \emptyset$.

\renewcommand{\theequation}{A-\arabic{equation}}
  \setcounter{equation}{0}  
  \section*{Appendix: Derivation of  mixing constraints}  

Neglecting $O(\epsilon^2)$ terms, we may write,
\begin{eqnarray}
|\psi^{\nu,l}_1\rangle_f &=& |\psi^{0 \nu,l}_1\rangle_f + 
                       \epsilon^{\nu,l}_{12}|\psi^{0 \nu,l}_2\rangle_f+
                       \epsilon^{\nu,l}_{13}|\psi^{0 \nu,l}_3\rangle_f, \\
|\psi^{\nu,l}_2\rangle_f &=& -{\epsilon^{\nu,l}_{12}}^*|\psi^{0 \nu,l}_1\rangle_f + 
                       |\psi^{0 \nu,l}_2\rangle_f+
                       \epsilon^{\nu,l}_{23}|\psi^{0 \nu,l}_3\rangle_f, \\
|\psi^{\nu,l}_3\rangle_f &=& -{\epsilon^{\nu,l}_{12}}^* |\psi^{0 \nu,l}_1\rangle_f - 
                      {\epsilon^{\nu,l}_{23}}^* |\psi^{0 \nu,l}_2\rangle_f+
                      |\psi^{0 \nu,l}_3\rangle_f ,
\end{eqnarray}
where
\begin{equation}
|\psi^{0 \nu}_1 \rangle_f=\begin{pmatrix}
\sqrt{2 \over 3} \cr -\sqrt{1 \over 6} \cr \sqrt{1 \over 6} \end{pmatrix},
|\psi^{0 \nu}_2 \rangle_f=\begin{pmatrix}\sqrt{1 \over 3} \cr \sqrt{1 \over 3} \cr -\sqrt{1 \over 3} \end{pmatrix},
|\psi^{0 \nu}_3 \rangle_f=\begin{pmatrix} 0 \cr \sqrt{1 \over 2} \cr \sqrt{1 \over 2} \end{pmatrix}
\end{equation}
and
\begin{equation}
|\psi^{0 l}_1 \rangle_f=\begin{pmatrix}1 \cr 0 \cr 0 \end{pmatrix},
|\psi^{0 l}_2 \rangle_f=\begin{pmatrix}0 \cr 1 \cr 0 \end{pmatrix},
|\psi^{0 l}_3 \rangle_f=\begin{pmatrix}0 \cr 0 \cr 1 \end{pmatrix}.
\end{equation}

By using $U_\ell=(|\psi^\ell_1\rangle_f~|\psi^\ell_2\rangle_f~|\psi^\ell_2\rangle_f)$
and
$U_\nu=(|\psi^\nu_1\rangle_f~|\psi^\nu_2\rangle_f~|\psi^\nu_3\rangle_f)$,
one is led to the respective expressions for $U_\ell$ and $U_\nu$, as given in the
text. If we define
$L \equiv U^\dagger_\ell U_\nu$, then neglecting
$O(\epsilon^2)$ terms, the nine elements of the $L$ matrix are
\begin{eqnarray}
L_{11} &=&
\sqrt{2 \over 3} 
+ \sqrt{1 \over 3} \epsilon^\nu_{12}
-\sqrt{1 \over 6} \epsilon^{\ell *}_{12}
+\sqrt{1 \over 6} \epsilon^{\ell *}_{13}, \\
L_{12} &=&
\sqrt{1 \over 3} 
- \sqrt{2 \over 3} \epsilon^{\nu *}_{12}
+\sqrt{1 \over 3} \epsilon^{\ell *}_{12}
-\sqrt{1 \over 3} \epsilon^{\ell *}_{13},\\
L_{13} &=&
-\sqrt{2 \over 3} \epsilon^{\nu *}_{13} 
-\sqrt{1 \over 3} {\epsilon^{\nu *}_{23}}
+\sqrt{1 \over 2} {\epsilon^{\ell *}_{12}}
+\sqrt{1 \over 2} {\epsilon^{\ell *}_{13}} ,\\
L_{21} &=&
-\sqrt{1 \over 6} 
+ \sqrt{1 \over 3} \epsilon^\nu_{12}
+\sqrt{1 \over 2} \epsilon^{\nu}_{13}
-\sqrt{2 \over 3} \epsilon^{\ell}_{12}
+\sqrt{1 \over 6} {\epsilon^{\ell *}_{23}},\\
L_{22} &=&
\sqrt{1 \over 3} 
+ \sqrt{1 \over 6} \epsilon^{\nu *}_{12}
+\sqrt{1 \over 2}  \epsilon^{\nu}_{23}
-\sqrt{1 \over 3}  \epsilon^{\ell}_{12}
-\sqrt{1 \over 3}  \epsilon^{\ell *}_{12}, \\
L_{23} &=&
\sqrt{1 \over 2} 
+ \sqrt{1 \over 6} \epsilon^{\nu *}_{13}
-\sqrt{1 \over 3} \epsilon^{\nu *}_{23}
+\sqrt{1 \over 2} \epsilon^{\ell *}_{23}, \\
L_{31} &=&
\sqrt{1 \over 6} 
- \sqrt{1 \over 3} \epsilon^\nu_{12}
+\sqrt{1 \over 2} {\epsilon^{\nu}_{13}}
-\sqrt{2 \over 3} {\epsilon^{\ell}_{13}}
+\sqrt{1 \over 6} \epsilon^\ell_{23}, \\
L_{32} &=&
-\sqrt{1 \over 3} 
- \sqrt{1 \over 6} {\epsilon^{\nu *}_{12}}
+\sqrt{1 \over 2} {\epsilon^{\nu}_{23}}
-\sqrt{1 \over 3} {\epsilon^{\ell}_{13}}
-\sqrt{1 \over 3} \epsilon^\ell_{23}, \\
L_{33} &=&
\sqrt{1 \over 2} 
- \sqrt{1 \over 6} {\epsilon^{\nu *}_{13}}
+ \sqrt{1 \over 3} {\epsilon^{\nu *}_{23}}
-\sqrt{1 \over 2} {\epsilon^{\ell }_{23}}. 
\end{eqnarray}
Similarly, defining $N \equiv U_{PMNS}K^{-1}$ and again neglecting $O(\epsilon^2)$ terms
\begin{equation}
N=\begin{pmatrix}c_{12} & s_{12} & s_{13}~e^{-i\delta_{CP}} \cr
          -s_{12}c_{23}-\sqrt{1 \over 3}s_{13}e^{i \delta_{CP}} 
           & c_{12}c_{23} -\sqrt{1 \over 6} s_{13}e^{i \delta_{CP}} & s_{23} \cr
s_{12}s_{23}-\sqrt{1 \over 3} s_{13} e^{i \delta_{CP}} & -c_{12}s_{23}-\sqrt{1 \over 6} s_{13}e^{i \delta_{CP}}
           & c_{23} \end{pmatrix}.
\end{equation}

Expanding in $\epsilon$, the relations $\sqrt{2}c_{12}+s_{12}=\sqrt{3}+O(\epsilon^2)$ and
$c_{23}+s_{23}=\sqrt{2} + O(\epsilon^2)$ are automatic. The equality $L=N$ leads to the mixing 
constraint relations. Specifically, the identification 
of elements or their combinations
\begin{eqnarray}
&& L_{11}=N_{11},~L_{21}-L_{31}=N_{21}-N_{31},L_{21}+L_{31}=N_{21}+N_{31},\nonumber\\
&& L_{12}=N_{12}, L_{22}+L_{32}=N_{22}+N_{32},~L_{22}-L_{32}=N_{22}-N_{32}, \nonumber\\
&& L^*_{13}=N^*_{13},~L_{33}+L_{23}=N_{33}+N_{23},~L_{33}-L_{23}=N_{33}-N_{23}, \nonumber
\end{eqnarray}
neglecting $O(\epsilon^2)$ terms, lead respectively to the equations 
\begin{eqnarray}
&& c_{12}-\sqrt{2 \over 3} = {1 \over \sqrt{2}}~(\sqrt{1 \over 3}-s_{12})
=\sqrt{1 \over 3} \epsilon^\nu_{12}-\sqrt{1 \over 6}({\epsilon^{\ell *}_{12}} -{\epsilon^{\ell *}_{13}} ),\label{a16}\\
&& -\sqrt{1 \over 3}(c_{23}-s_{23})-{2 \over \sqrt{3}}s_{13}e^{i \delta_{CP}} = \sqrt{2} 
\epsilon^\nu_{13}-\sqrt{2 \over 3}(\epsilon^\ell_{12}+\epsilon^\ell_{13})
+\sqrt{1 \over 6}(\epsilon^\ell_{23}+ {\epsilon^{\ell *}_{23}}),\label{a17}\\
&&-\sqrt{2}s_{12} = -2 \sqrt{1 \over 6}+2\sqrt{1 \over 3}\epsilon^\nu_{12}
-\sqrt{2 \over 3}(\epsilon^\ell_{12}-\epsilon^\ell_{13})
+\sqrt{1 \over 6}(\epsilon^\ell_{23}-{\epsilon^{\ell *}_{23}}),\label{a18}\\
&& s_{12}= \sqrt{1 \over 3}-\sqrt{2 \over 3}{\epsilon^{\nu *}_{12}}
+\sqrt{1 \over 3}({\epsilon^{\ell *}_{12}}-{\epsilon^{\ell *}_{13}}),\label{a19}\\
&&\sqrt{2 \over 3}\left( c_{23} -s_{23} -s_{13} e^{i \delta_{CP}}\right)
= \sqrt{2}\epsilon^\nu_{23}-\sqrt{1 \over 3}
(\epsilon^\ell_{12}+ \epsilon^\ell_{13}+{\epsilon^{\ell *}_{23}} + \epsilon^\ell_{23}),\label{a20}\\
&& \sqrt{2}c_{12}=2 \sqrt{1 \over 3} + 2 \sqrt{1 \over 6}\epsilon^{\nu *}_{12}
-\sqrt{1 \over 3}({\epsilon^{\ell *}_{23}}-\epsilon^\ell_{23})
-\sqrt{1 \over 3}(\epsilon^\ell_{12}-\epsilon^\ell_{13}),\label{a21}\\
&& s_{13}e^{i \delta_{CP}}=-\sqrt{1 \over 3}(\sqrt{2} \epsilon^\nu_{13} + \epsilon^\nu_{23})
+\sqrt{1 \over 2}(\epsilon^\ell_{12}+ \epsilon^\ell_{13})
\label{a22}\\
&&c_{23}+s_{23}=\sqrt{2}+{1 \over \sqrt{2}}({\epsilon^{\ell *}_{23}}-\epsilon^\ell_{23}),
\label{a23}\\
&&c_{23}-s_{23}=-\sqrt{2 \over 3}{\epsilon^{\nu *}_{13}}
-\sqrt{1 \over 2}(\epsilon^\ell_{23}+\epsilon^{\ell *}_{23})
+ {2 \over \sqrt{3}} \epsilon^{\nu *}_{23}. \label{a24}
\end{eqnarray}

Eq. (\ref{eqn-27}) is a direct conseqence of (\ref{a23}). Eq. (\ref{eqn-26}) is easily derived from
(\ref{a20}) and (\ref{a22}), while Eq. (\ref{eqn-25}) follows from (\ref{a19}) and (\ref{a21}).
Now Eq. (\ref{eqn-28}) obtains from (\ref{a21}), whereas Eq. (\ref{eqn-29}) is just a rewritten 
form of (\ref{a16}) with the input of Eq. (\ref{eqn-28}). Eq. (\ref{eqn-30}) follows from (\ref{a17}) and (\ref{a20}). 
Finally, Eq. (\ref{eqn-31}) is the same as (\ref{a22}).

\section*{ Acknowledgement} 
A part of this work was done at the WHEPP13 workshop, Puri, India. 
We thank A. Dighe, P. Byakti, S. Choubey, A. Raychaudhuri 
and S. Uma Sankar for their comments. PR
acknowledges a Senior Scientistship of Indian National Science Academy.


\begin{thebibliography}{100} 

\bibitem{nf-1} J. Beringer et. al. (PDG), Phys. Rev. {\bf D86}, 010001 (2012).


\bibitem{nf-2}
R. N. Mohapatra 
and A. Y. Smirnov, Ann. Rev Nucl Part. Sci {\bf 56}, 569 (2006). 



\bibitem{nf-3a} 
A. de Gouvea et al, arXiv:1310.4340 [hep-ph].

\bibitem{nf-3b}
 S. Parke, arXiv:1310.5992 [hep-ph].

\bibitem{nf-3c} 
H. Minakata, arXiv:1403.3276 [hep-ph].




\bibitem{nf-4} 
J. Lesgourgues and S. Pastor, Adv. High 
Energy Phys. {\bf 2012}, 608515 (2012).



\bibitem{nf-5a}
G. Atlarelli and F. Feruglio, New J. Phys. {\bf 6}, 106 (2006).

\bibitem{nf-5b}
 G. Altarelli, 
F. Feruglio, L. Merlo and E. Stamou, JHEP {\bf 1208}, 012 (2012).

\bibitem{nf-5c}
 S. King 
and C. Luhn, Rept. Prog. Phys. {\bf T6}, 006201 (2013).




\bibitem{nf-6} 
P. F. Harrison, D. H. Perkins and 
W. G. Scott, Phys. Lett. {\bf B530}, 167 (2002).




\bibitem{nf-7} 
G. Altarelli, S. Feruglio and L. Merlo, Fortsch. Phys. {\bf 61}, 507 (2013).


\bibitem{nf-8na}
S-F. Ge, D. A. Dicus, W. W. Repko, Phys. Lett. {\bf B702} 220, (2011).


\bibitem{nf-8nb}
 S. F. King 
and C. Luhn, JHEP 11099, 042 (2011). 



\bibitem{nf-8a}
S-F. Ge, D. A. Dicus, W. W. Repko, Phys. Rev. Lett. {\bf 108} 041801, (2012).


\bibitem{nf-8b} 
D. Hernandez and A. Y Smirnov, Phys. Rev. {\bf D 86}, 
053014 (2012); {\it ibid} {\bf D87}, 053005 (2013).


\bibitem{nf-8c}
G. Altarelli, S. 
Feruglio, L. Merlo and E. Stamou, {\it loc.cit}.


\bibitem{nf-8nc}
A. D. Hanlon, S-F. Ge, W. W. Repko, Phys. Lett. {\bf 729}, 185 (2014).



\bibitem{nf-9} 
M. C. Gonzalez-Garcia, M. Maltoni, J. Salvado and 
T. Schwetz, JHEP {\bf 1212}, 123 (2012). 



\bibitem{nf-10} 
D. V. Forero, M. Tortola 
and J. W. F. Valle, Phys. Rev. {\bf D86}, 073012 (2012).



\bibitem{nf-11}
F. Capozzi, G. L. Fogli, E. Lisi, A. Marrone, D. Montanino and A. Palazzo, 
Phys. Rev. {\bf D89}, 093018 (2014).


\bibitem{nf-12a}
F. P. An et al, Phys. Rev. Lett. {\bf 108}, 
171803 (2012).

\bibitem{nf-12b}
 J. K. Ahn et.al, Phys. Rev. Lett. {\bf 108}, 
191802 (2012).


\bibitem{nf-12c}
 K. Abe et al; Phys. Rev. Lett. {\bf 107}, 
181802 (2011).


\bibitem{nf-12d}
 Y. Abe et al, Phys. Rev. Lett. {\bf 108}, 131801 (2012).


\bibitem{nf-13new}
Z.Z. Xing, Phys. Lett. {\bf B 533}, 85 (2002).


\bibitem{nf-13a}
X. He 
and A. Zee, Phys. Rev. {\bf D84}, 053004 (2011).



\bibitem{nf-13b}  
D. A. Sierra, I. de M. Varzielas 
and E. Houet, Phys. Rev. {\bf D87}, 093009 (2013).

\bibitem{nf-13c} 
D. Borah, Nucl. Phys. {\bf B876}, 575 (2013).


\bibitem{nf-new-1} K. Bora, D. Dutta and P. Ghoshal, arXiv:1405.7182 [hep-ph]


\bibitem{nf-new-2} Soumya C., K. N. Deepthi and R. Mohanta, arXiv:1408.6071 [hep-ph]


\bibitem{nf-14a} B. Brahmachari and 
A. Raychaudhuri, Phys. Rev. {\bf D86}, {\bf R}051302 (2012).


\bibitem{nf-14b}
L. J. Hall and G. G. Ross, JHEP {\bf 1311}, 091 (2013).


\bibitem{nf-14c} 
S. Pramanick and A. Raychaudhuri, Phys. Rev. {\bf D88}, 093009 (2013).



\bibitem{nf-15a} S. K. Agarwalla, S. Prakash and S. Uma Sankar, JHEP {\bf 1307}, 131 (2013).


\bibitem{nf-15b} 
A. Chatterjee, P. Ghoshal, S. Goswami and S. K. Raut, JHEP {\bf 1306}, 010 (2013)


\bibitem{nf-15c} M. Ghosh, P. Ghoshal, S. Goswami, and S. K. Raut, Nucl. Phys. {\bf B884}, 274 (2014).


\bibitem{nf-16} P. A. N. Machado, H. Minakata, H. Nunokawa and R. R. Funchal, 
JHEP {\bf 1405}, 109 (2014).

\bibitem{nf-17} C. Adams et al, arXiv:1307.7335 [hep-ph].

\bibitem{nf-18a} A. Stahl et al, Report No. CERN-SPSC-2012-021.


\bibitem{nf-18b}
 S. K. Agarwalla et al. JHEP {\bf 1405}, 094 (2014)

\bibitem{nf-19} International Design Study of the Neutrino Factory, http://www.ids-nf.org


\bibitem{nf-39} G. Altarelli and F. Feruglio, Nucl. Phys. {\bf B720}, 64 (2005).

\bibitem{nf-40} M. Honda and M. Tanimoto, Prog. Theor. Phys. {\bf 119}, 583 (2008).

\bibitem{nf-41} B. Brahmachari, S. Choubey and M. Mitra, Phys. Rev. {\bf D77}, 073008 (2008).

\bibitem{nf-42}G. Alterelli, F. Feruglio, L. Merlo and E. Stamou, JHEP {\bf 1208}, 021 (2012).


\bibitem{nf-43}J. Berry and W. Rodejohann, Phys. Rev. {\bf D81}, 093002(2010); 
errtm. $ibid$, {\bf D81}, 119901 (2010).


\bibitem{nf-44}S. Goswami, S. T. Petcov, S. Ray 
and W. Rodejohann, Phys. Rev. {\bf D80}, 053013 (2009).


\end{thebibliography}
\end{document}